\documentclass[english]{article}
\usepackage[T1]{fontenc}
\usepackage[latin9]{inputenc}
\usepackage{units}
\usepackage{amsmath}
\usepackage{graphicx}
\usepackage{amssymb}
\usepackage{babel}
\usepackage{nopageno}
\usepackage{url}
\usepackage{epstopdf}

\title{Surface Curvature Effects on Reflectance from Translucent Materials\footnote{The first version of this paper was published in the Communication Papers Proceedings of 18th International Conference on Computer Graphics, Visualization and Computer Vision 2010 - WSCG2010, pp. 169-172.}}

\author{
\parbox{\textwidth}{\centering
Konstantin Kolchin\\
\small
NVIDIA Corporation, Russian branch\\
Arbat 10, Moscow 119002, Russia \\
{\small{} kkolchin@nvidia.com }
}
}

\begin{document}
\maketitle

\begin{abstract}
Most of the physically based techniques for rendering translucent
objects use the diffusion theory of light scattering in turbid media.
The widely used dipole diffusion model \cite{Jensen2001} applies
the diffusion-theory formula derived for the planar surface to objects
of arbitrary shapes. This paper presents first results of 
our investigation of how surface
curvature affects the diffuse reflectance from translucent materials. 
\end{abstract}

\section{Introduction}

Translucent materials, such as human skin, marble, wax, fruits, more
scatter light than absorb it. Therefore, when a photon enters such
a material, it undergoes many scattering events under the surface
before it leaves the material. Such a light behavior is well described
by the Bidirectional Surface Scattering Distribution Function (BSSRDF)
\cite{Nicodemus}. Based on the light diffusion theory, Jensen et
al. \cite{Jensen2001} suggested the dipole diffusion model for BSSRDF.
This model applies an expression for reflectance from a turbid half-space
to arbitrarily shaped objects. The multipole \cite{Multipole,JOSA}
and quadpole \cite{Jensen2008} models have been suggested to describe
more complicated geometries - a multilayered slab (or half-space)
and a right-angle corner, respectively. Jensen et al. \cite{Jensen2008}
showed that a big variety of shapes can be rendered by combining photon
tracing and a scheme for interpolating between dipole and quadpole
and between quadpole and multipole models wherever appropriate. However,
they do not focus on how the BSSDRF itself changes as a flat surface
is replaced with a curved one. It is difficult to devise how their
interpolation scheme can be used with approaches that do not use photon
tracing - for example, the curvature-based method \cite{Kolchin}.
Our goal is to investigate how inclusion of curvature may change the
diffusion BSSRDF model. A BSSRDF model that includes curvature effects
could be easily incorporated into many existing approaches for rendering
translucent materials. We present here preliminary results of our
study.

\section{Diffusion Equation}

Under the assumption that light scattering in a turbid medium dominates
absorption, light transport in it is well described with the diffusion
theory \cite{Farrell1992}. The fluence rate $\Psi(\mathbf{r})$ obeys
the modified Helmholtz equation \cite{Farrell1992}

\begin{equation}
\Delta\Psi-\sigma_{tr}^{2}\Psi=-D^{-1}\delta(\mathbf{r}-\mathbf{r}_{0})\label{eq:HelmEq}\end{equation}
 where $\sigma_{tr}=\sqrt{3\sigma_{a}(\sigma_{s}'+\sigma_{a})}$ is
the effective transport coefficient, $\sigma_{s}'$ is the reduced
scattering coefficient, $\sigma_{a}$ is the absorption coefficient,
$D=\frac{1}{3(\sigma_{s}'+\sigma_{a})}$ is the diffusion coefficient.
We refer the reader to \cite{Jensen2001,Farrell1992} for explanation
of the physical meaning of the quanitities. In the above equation,
we assume that there is a single source in the medium, and it is located
at a point $\mathbf{r}_{0}$.

Let us first consider the case of translucent material occupying the
half-space $z>0$. The point source is at $\mathbf{r}_{0}=(0,0,z_{0})$.
Farrell et al. \cite{Farrell1992} showed that quite an accurate solution
can be obtained by using the boundary condition $\Psi|_{z=-z_{b}}=0$
and putting the image source at the point $\mathbf{r}_{0}=(0,0,-z_{0}-2z_{b})$,
where $z_{b}=2AD$, and $A$ is calculated as described in \cite{Jensen2001,Farrell1992}.
The resulting fluence is

\[
\Psi(\rho,z_{0})=\frac{1}{4\pi D}\left(\frac{e^{-\sigma_{tr}r_{1}}}{r_{1}}+\frac{e^{-\sigma_{tr}r_{2}}}{r_{2}}\right),\]
 where $r_{1}$ and $r_{2}$ are the distances to the source and image
source, respectively; that is,

\begin{eqnarray}
r_{1} & = & [(z-z_{0})^{2}+\rho^{2}]^{1/2}\\
r_{2} & = & [(z+z_{0}+2z_{b})^{2}+\rho^{2}]^{1/2}\end{eqnarray}

The reflectance is calculated from the fluence using the formula

\begin{equation}
R=-D\nabla\Psi\label{eq:Flux}\end{equation}
 where the gradient is evaluated at the interface. In the planar case,
this gives

\begin{eqnarray}
R(\rho,z_{0}) & = & \frac{1}{4\pi}\left[ z_{0}(\sigma_{tr}+\frac{1}{r_{1}})\frac{e^{-\sigma_{tr}r_{1}}}{r_{1}^{2}}+\right.\nonumber \\
 & \quad & \left.+(z_{0}+2z_{b})(\sigma_{tr}+\frac{1}{r_{2}^{2}})\frac{e^{-\sigma_{tr}r_{2}}}{r_{2}^{2}}\right]\end{eqnarray}
 where $r_{1}$ and $r_{2}$ are calculated for $z=0$.

The dipole diffusion model \cite{Jensen2001} applies the above formula
to an arbitrary shaped air-material interface by calculating $r_{1}$
and $r_{2}$ as the distance from a point being shaded to the source
and image source, respectively.

\section{Exact Solution for a Sphere}

\label{sec:Sphere}

Suppose the turbid medium is confined within a sphere having the radius
$R_{0}$ and the center at $z=R_{0}$. In addition to Cartesian coordinates,
we will also use the polar system of coordinates with $r$ counted
from the sphere center and $\theta$counted from the $z$ axis. We
assume that $R_{0}$is much bigger than the mean free path for photons
scattered in the medium, we can use the same boundary condition as
in the planar case - namely, the fluence rate vanishes at a distance
of $z_{b}$ from the sphere surface. In other words, $\Psi$ is zero
at a sphere of the radius $R=R_{0}+z_{b}$. We will solve eq. (\ref{eq:HelmEq})
with the boundary condition $\Psi|_{r=R}=0$ following the method
described in \cite{MathMethods}. The solution of the modified Helmholtz
equation \ref{eq:HelmEq}with the zero boundary condition on the sphere
$r=R$ can be written as

\begin{eqnarray}
\Psi(r,\theta)=\begin{cases}
\sum_{m=0}^{\infty}A_{m}\frac{I_{m+1/2}(\sigma_{tr}r)}{\sqrt{r}}P_{m}(\cos\theta)\mbox{ , }r<r'\\
\sum_{m=0}^{\infty}B_{m}\frac{1}{\sqrt{r}}[I_{m+1/2}(\sigma_{tr}r)\times\\
\times K_{m+1/2}(\sigma_{tr}R)-K_{m+1/2}(\sigma_{tr}r)\times\\
\times I(\sigma_{tr}R)]P_{m}(\cos\theta)\mbox{ , }r>r'\end{cases}\label{eq:GenSolution}\end{eqnarray}
 where $r'$ is the distance of the point source from the sphere center;
that is, we suppose that$\mathbf{r}_{0}$ has the polar coordinates
$r=r'$ and $\theta=0$. The functions $I_{v}(r)$ and $K_{\nu}(r)$
are the modified Bessel functions \cite{Abramovitz}. The constants
$A_{m}$and $B_{m}$ are determined by stitching the solutions \ref{eq:GenSolution}
at the sphere $r=r'$. The function $\Psi$ is continuous, but its
derivative is not. In a manner similar to that used in \cite{MathMethods},
we integrate eq. \ref{eq:HelmEq} over an infinitisemally thin region
confined by parts of spherical surfaces with radiuses $r=r'+\epsilon$
and $r=r'-\epsilon$ and containing the point$\mathbf{r}_{0}$. We
utilize the Gauss theorem and get

\begin{equation}
\left(\frac{\partial\Psi}{\partial r}\Big|_{r'+\epsilon}-\frac{\partial\Psi}{\partial r}\Big|_{r'-\epsilon}\right)=\frac{1}{r'^{2}}\delta(\Omega)\label{eq:Jump}\end{equation}
 where $\Omega$ is the solid angle variable. The delta function $\delta(\Omega)$
can be decomposed in terms of the Legendre polynomials as \cite{MathMethods}

\begin{equation}
\delta(\Omega)=\sum_{m=0}^{\infty}\frac{(2m+1)}{4\pi}P_{m}(\cos\theta)\label{eq:DeltaDecomposition}\end{equation}

Substituting eq. (\ref{eq:DeltaDecomposition}) into eq. (\ref{eq:Jump})
and calculating the derivatives from eq. (\ref{eq:GenSolution}),
we arrive at an equation for $A_{m}$and $B_{m}$. One more equation
for them is obtained by requiring continuity of $\Psi$ at $r=r'$.
Solving the resulting system of two equations, we get

\begin{eqnarray*}
\Psi(r,\theta) & = & \frac{1}{4\pi D}\left[\sum_{m=0}^{\infty}\frac{(2m+1)}{\sqrt{rr'}}I_{m+1/2}(\sigma_{tr}r')\times \right.\\
 & \quad & \left. \times I_{m+1/2}(\sigma_{tr}r)\frac{K_{m+1/2}(\sigma_{tr}R)}{I_{m+1/2}(\sigma_{tr}R)}P_{m}(\cos\theta)-\frac{e^{-\sigma_{tr}\widetilde{r}}}{\widetilde{r}}\right]\end{eqnarray*}
 where

\[
\tilde{r}=(r^{2}+r'^{2}-2rr'\cos\theta)^{\nicefrac{1}{2}},\]
 and we used equality 10.2.35 from \cite{Abramovitz}.

To find the reflectance, we choose $r'=R_{0}-z_{0}$, apply eq. \ref{eq:Flux}
and set $r=R_{0}$ and get

\begin{eqnarray*}
R(r,\theta) & = & \frac{1}{4\pi}\sigma_{tr}\left\{\sum_{m=0}^{\infty}\frac{(2m+1)}{\sqrt{R_{0}r'}}I_{m+1/2}(\sigma_{tr}r')\times \right.\\
 &  & \times I'_{m+1/2}(\sigma_{tr}R_{0})\frac{K_{m+1/2}(\sigma_{tr}R)}{I_{m+1/2}(\sigma_{tr}R)}P_{m}(\cos\theta)+\\
 &  & \left.+[z_{0}\cos\theta-R_{0}(\cos\theta-1)](\sigma_{tr}+\frac{1}{r_{1}})\frac{e^{-\sigma_{tr}r_{1}}}{r_{1}^{2}}\right\}\end{eqnarray*}

\section{Results}

\begin{figure}[htb]
 \includegraphics[width=1.05\linewidth]{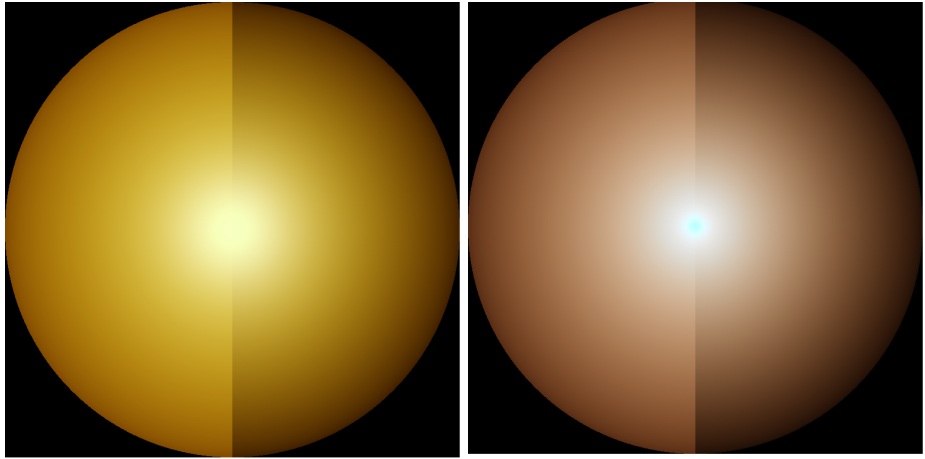} 

\caption{\label{fig:ex1} A spherical potato (left) and a marble sphere (right)
illuminated with a stencil beam, which enters at the image center,
normally to the image plane. Each of the spheres is rendered using
the exact solution proposed (left part of a sphere) and the dipole
diffusion model (right part of a sphere).}

\end{figure}

We calculated the reflectance from translucent spheres of various
radiuses. The incident light is a pencil beam entering a sphere at
$x=0,$y=0 . Ideally, we should consider a line of sources situated
along the $z$ axis. But it was shown in \cite{Farrell1992} that
they all can be replaced with a single source located $z=\nicefrac{1}{(\sigma_{s}'+\sigma_{a})}$.
The plot below shows how the reflectance depends on the distance from
the point of light entrance measured along the surface (that is, the
length of a geodesic connecting the entrance point and the point of
interest). The calculations were done for the scattering coefficient
$\sigma_{s}'=1mm^{-1}$and absorption coefficient $\sigma_{a}=0.01mm^{-1}$
(note that in \cite{Farrell1992}, the same quantities are designated
as $\mu_{s}'$ and $\mu_{a}$, respectively). These values of the
scattering and absorption coefficients are typical for human tissue
(see \cite{Jensen2001}). It can be seen that in this case, the difference
between the exactly computed reflectance and that found by the dipole
diffusion model becomes noticable only when the radius approaches
1 cm.

Figure 1 above shows visualization of light reflection from spheres
having a radius of 1 cm in two cases - a potato, on the left, and
marble, on the right. As for the plot given below, we assume that
a sphere is lit up by a stencil beam entering the sphere at the center
of the image. The left part of each of the image corresponds to the
exact calculation we describe above. The right part is computed using
the diffuse dipole approximation. We used the measured values $\sigma_{s}'$
and $\sigma_{a}$ reported in \cite{Jensen2001}. Because the amount
of reflected light decays with distance from the entrance point very
rapidly, we applied the tone mapping operator to a calculated HDR
image. We chose the logarithmic mapping operator\cite{Drago2003},
as it is simple and robust, and a source code for its implementation
is available on the web.

Probably, we could anticipate in advance that the diffuse dipole model
would underestimate the reflectance. However, our investigation shows
that this underestimation is small when curvature radiuses are of
the scale of several centimeters and more for such materials as marble,
potato, human tissue.

\label{ReflPlot}\includegraphics[scale=0.57]{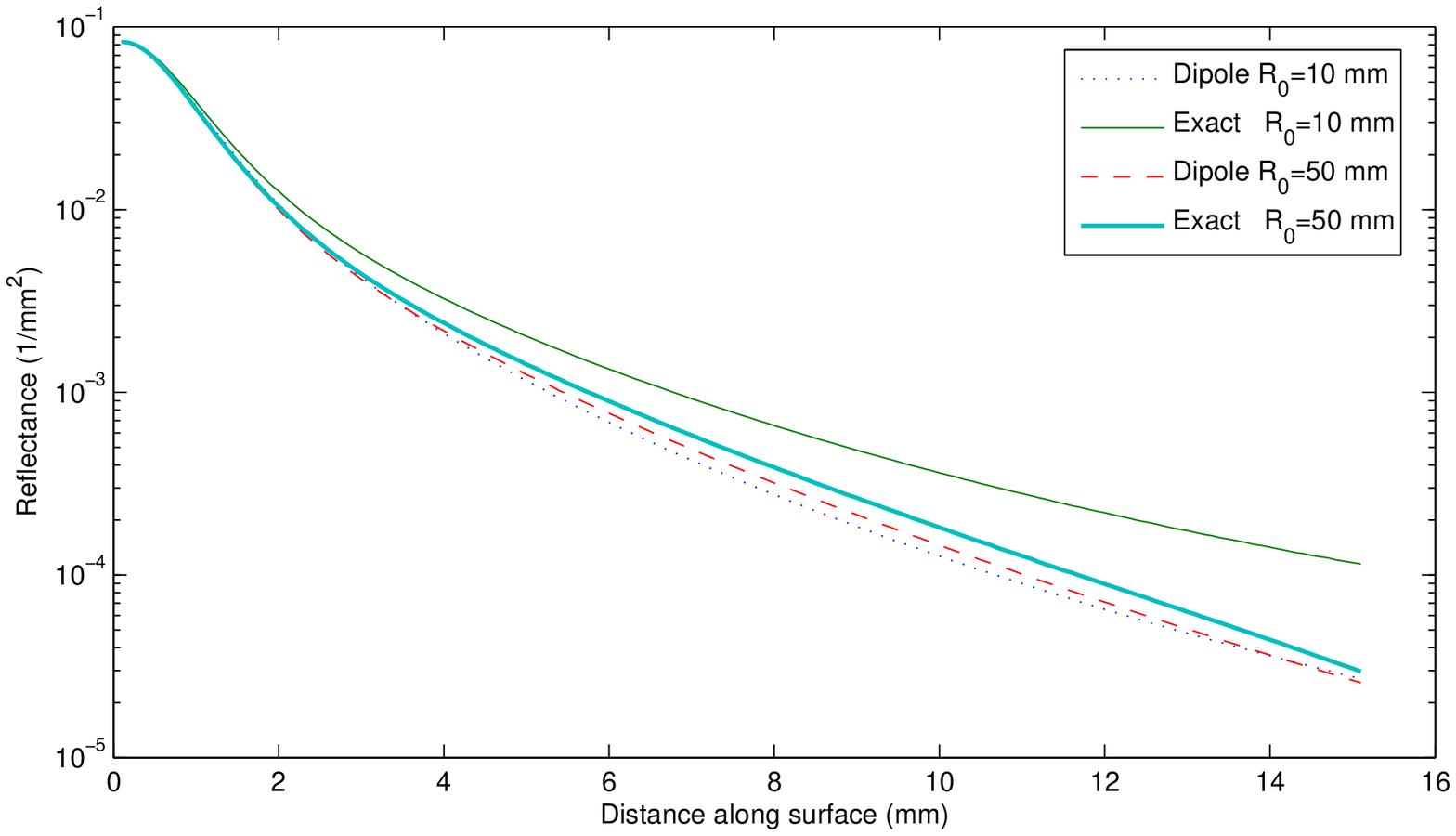}

\section{Future Work}

The investigation presented here definitely lacks comparison of analytical
results with Monte-Carlo simulations. We are working on this and plan
to report them elsewhere when the work is complete. Also, we would
like to consider the case of arbatrarily curved surfaces. It would
be interesting to try to build a phenomenological model for reflectance
from a translucent material with an arbitrary surface. It can be sought
as a function of principal curvatures at the point of light entrance.
An approximate solution for slightly curved surfaces (Appendix A presents
an approximate solution for the modified Helmholtz equation in the
case of a curved boundary) can serve as a base in attempts to construct
a phenomenological model. Monte-Carlo simulations can be used for
validation of such a model. A big potential of the phenomenological
approach to constructing BSSRDF models has been proven by successfull
development of an empirical BSSRDF model described in \cite{Jensen2009}.
A BSSRDF model including surface curvature could be incorporated into
the curvature-based method \cite{Kolchin}. It could be used for investigating
perceptional effects, such as color shift at the terminator line \cite{Green}.

\bibliographystyle{alpha} \bibliographystyle{alpha} \bibliographystyle{alpha}
\bibliography{te}

\appendix

\section{An Approximate Solution for Slightly Curved Surfaces}

We consider the modified Helmholtz equation

\begin{equation}
\Delta\Psi-\sigma_{tr}^{2}\Psi=-D^{-1}\delta(\mathbf{r}-\mathbf{r}_{0})\label{eq:HelmEq}\end{equation}
 where $\mathbf{r}$ is a point in three-dimensional space, and $\mathbf{r}=(x,y,z)$
in the Cartesian coordinate system. The source is located at the point
$\mathbf{r}_{0}=(0,0,a)$. We are searching for a solution of this
equation for $z>f(x,y)$ with the boundary condition

\[
\Psi|_{z=f(x,y)}=0\]
 where $z=f(x,y)$ is the equation describing some surface. Let's
assume deviation of the surface from a plane to be small and introduce
a small parameter, $\epsilon,$ to describe this, so that the surface
equation becomes

\[
z=\epsilon f(x,y),\]

Let us search a solution to the modified Helmhotz equation as a series
in $\epsilon$.

\[
\Psi=\Psi_{0}+\epsilon\Psi_{1}+\epsilon^{2}\Psi_{2}+...\]

Then $\Psi_{0}$ is the solution for the planar case, which is given
by

\[
\Psi_{0}=\frac{1}{4\pi D}\left(\frac{e^{-\sigma_{tr}r_{1}}}{r_{1}}+\frac{e^{-\sigma_{tr}r_{2}}}{r_{2}}\right),\]
 where $r_{1}$ and $r_{2}$ are the distances to the source and image
source, respectively; that is,

\begin{eqnarray}
r_{1} & = & [(z-a)^{2}+x^{2}+y^{2}]^{1/2}\\
r_{2} & = & [(z+a)^{2}+x^{2}+y^{2}]^{1/2}\end{eqnarray}

We know the Green function for the planar case, so let's try to find
the boundary condition for $\Psi_{1}$ at the plane $z=0$. It can
be found by extending $\Psi_{0}$ in $z$. We know that $\Psi$ takes
a zero value at the surface; therefore, we have

\[
\Psi|_{z=\epsilon f(x,y)}=0=\Psi_{0}|_{z=0}+\frac{\partial\Psi_{0}}{\partial z}\Big|_{z=0}\epsilon f(x,y)+\epsilon\Psi_{1}|_{z=0}+...\]

Each term in the $\epsilon$ series should be equal to zero; therefore,
we get

\begin{equation}
\Psi_{1}|_{z=0}=-\frac{\partial\Psi_{0}}{\partial z}\Big|_{z=0}f(x,y)\end{equation}

Therefore, $\Psi_{1}$for $z>0$ can be found as

\begin{equation}
\Psi_{1}(x,y,z)=\iint\frac{\partial\Psi_{0}(x',y',z')}{\partial z'}\Big|_{z'=0}f(x',y')\frac{\partial G}{\partial z'}\Big|_{z'=0}dx'dy'\label{eq:MainInt}\end{equation}
 where $G$ is the Green function for the Helmholtz equation and boundary
conditions of the form $\Psi|_{z=0}=\psi(x,y)$. Namely,

\[
G(x,y,z,x',y',z')=\frac{1}{4\pi D}\left(\frac{e^{-\sigma_{tr}r_{1}'}}{r_{1}'}+\frac{e^{-\sigma_{tr}r_{2}'}}{r_{2}'}\right),\]
 where $r_{1}$ and $r_{2}$ are the distances to the source, $(x,y,z),$
and the image source, $(x,y,-z),$ respectively; that is,

\begin{eqnarray}
r_{1}'=[(x'-x)^{2}+(y'-y)^{2}+(z'-z)^{2}]^{1/2}\\
r_{2}'=[(x'-x)^{2}+(y'-y)^{2}+(z'+z)^{2}]^{1/2}\end{eqnarray}
 We will calculate the integral in eq. \ref{eq:MainInt} using the
Fourier transform. Repeating the reasoning of \cite{Feinberg}, we
find that

\begin{equation}
\frac{\partial G}{\partial z'}\Big|_{z'=0}=\frac{1}{4\pi^{2}}\intop_{-\infty}^{+\infty}\intop_{-\infty}^{+\infty}e^{iq_{x}(x-x')+iq_{y}(y-y')-z\sqrt{q_{x}^{2}+q_{y}^{2}+\mu^{2}}}dq_{x}dq_{y}\end{equation}
 and

\begin{equation}
\frac{\partial\Psi_{0}(x',y',z')}{\partial z'}\Big|_{z'=0}=\frac{1}{4\pi^{2}D}\intop_{-\infty}^{+\infty}\intop_{-\infty}^{+\infty}e^{iq_{x}x'+iq_{y}y'-a\sqrt{q_{x}^{2}+q_{y}^{2}+\mu^{2}}}dq_{x}dq_{y}\label{eq:DPsi0}\end{equation}

Without loss of generality we may suppose that that the entrance point
is $(0,0,0)$, and the $(x,y)$ plane is tangent to the object surface
at the entrance point (this can always be achieved by a linear change
of space coordinates). We assume that the interface is a smooth surface.
It can then be locally represented as \[
z=f(x,y),\]
 where $f(x,y)$ is a smooth function, and its Taylor series expansion
begins from quadratic terms. Choosing the $x$ and $y$ axes along
the principal curvature directions, we get the following local surface
representation

\begin{equation}
f\left(x,y\right)=\frac{1}{2}\left(k_{1}x^{2}+k_{2}y^{2}\right)\label{eq:Monge}\end{equation}

\noindent where $k_{1}$,$k_{2}$ are the principal curvatures. The
above equation is accurate to the second order in $x$, $y$.

After substituting \ref{eq:Monge} into \ref{eq:MainInt}, we come
to expressions $k_{1}x'^{2}\frac{\partial\Psi_{0}(x',y',z')}{\partial z'} \big |_{z'=0}$
and $k_{2}y'^{2}\frac{\partial\Psi_{0}(x',y',z')}{\partial z'} \big |_{z'=0}$
inside the integral. Their Fourier transforms can be calculated by
taking the second derivative of the Fourier image of $\frac{\partial\Psi_{0}(x',y',z')}{\partial z'}|_{z'=0}$
in \ref{eq:DPsi0} with respect to $q_{x}$ and $q_{y}$, rescpectively.
Thus, the contribution to $\Psi_{1}$ coming from the $x$ coordinate
is (the $y$ part can be obtained by replacing $x$ with $y$)

\begin{eqnarray}
\Psi_{1} & = & \frac{k_{1}a}{4\pi^{2}D}\intop_{-\infty}^{+\infty}\intop_{-\infty}^{+\infty}\left\{\frac{1}{\sqrt{q_{x}^{2}+q_{y}^{2}+\mu^{2}}}-\left[\frac{q_{x}^{2}}{(q_{x}^{2}+q_{y}^{2}+\mu^{2})^{3/2}}+\frac{aq_{x}^{2}}{q_{x}^{2}+q_{y}^{2}+\mu^{2}}\right]\right\}\times\nonumber \\
 & \quad & \times e^{iq_{x}x+iq_{y}y-(z+a)\sqrt{q_{x}^{2}+q_{y}^{2}+\mu^{2}}}dq_{x}dq_{y}\end{eqnarray}

Integration by parts gives

\begin{eqnarray*}
\Psi_{1} & = & \frac{k_{1}a}{4\pi^{2}D}\intop_{-\infty}^{+\infty}\intop_{-\infty}^{+\infty}\left[\frac{(-iq_{x}x)}{\sqrt{q_{x}^{2}+q_{y}^{2}+\mu^{2}}}+\frac{zq_{x}^{2}}{q_{x}^{2}+q_{y}^{2}+\mu^{2}}\right] \times \\
 &  & \times e^{iq_{x}x+iq_{y}y-(z+a)\sqrt{q_{x}^{2}+q_{y}^{2}+\mu^{2}}}dq_{x}dq_{y}\end{eqnarray*}

Taking into account that

\begin{equation}
\frac{e^{-\sigma_{tr}r_{2}}}{r_{2}}=\frac{1}{2\pi}\intop_{-\infty}^{+\infty}\intop_{-\infty}^{+\infty}\frac{e^{iq_{x}x+iq_{y}y-(z+a)\sqrt{q_{x}^{2}+q_{y}^{2}+\mu^{2}}}}{\sqrt{q_{x}^{2}+q_{y}^{2}+\mu^{2}}}dq_{x}dq_{y}\end{equation}
 and

\begin{equation}
J_{0}(qr)=\frac{1}{2\pi}\intop_{-\pi}^{\pi}e^{iqr\cos\phi}d\phi\end{equation}
 we arrive at

\begin{equation}
\Psi_{1}=-\frac{k_{1}a}{2\pi D}\left[ x\frac{\partial}{\partial x}\left(\frac{e^{-\sigma_{tr}r_{2}}}{r_{2}}\right)+\frac{\partial^{2}}{\partial x^{2}}I(x,y,z)\right]\end{equation}
 where

\begin{equation}
I(x,y,z)=\intop_{0}^{+\infty}\frac{J_{0}(q\sqrt{x^{2}+y^{2}})e^{-(z+a)\sqrt{q^{2}+\mu^{2}}}}{q^{2}+\mu^{2}}qdq\end{equation}

Adding the contribution coming from the $y$ coordinate,
we get

\begin{equation}
\Psi_{1}=-\frac{a}{2\pi D}\left[(k_{1}x\frac{\partial}{\partial x}+k_{2}y\frac{\partial}{\partial y})(\frac{e^{-\sigma_{tr}r_{2}}}{r_{2}})+(k_{1}\frac{\partial^{2}}{\partial x^{2}}+k_{2}\frac{\partial^{2}}{\partial y^{2}})I(x,y,z)\right]\end{equation}

\end{document}